\documentclass[graphicx, nofootinbib, notitlepage, twocolumn, amsmath, amssymb, longbibliography]{revtex4-2}
\usepackage{dcolumn}  
\usepackage{bm}       
\usepackage{braket}
\usepackage{dsfont}
\usepackage{enumitem}
\usepackage{graphicx}
\usepackage{caption}
\usepackage{subcaption}
\usepackage[pdftex,bookmarks,breaklinks,bookmarksopen,bookmarksnumbered,linktocpage,pdftex]{hyperref}

\hypersetup{
            citecolor=red,  
            linkcolor=black,  
            urlcolor=blue }  

\def\nonum{\nonumber \\}
\def\mbf{\mathbf}
\def\t{\text}

\def\beit{\begin{itemize}}
\def\eit{\end{itemize}}
\def\mbf{\mathbf}

\def\l{\left}
\def\r{\right}

\def\kb{\text{k}_\text{B}}

\begin{document}
\title{Chemical equilibrium under vibrational strong coupling}
\author{Kaihong Sun}
\author{Raphael F. Ribeiro}
\email[]{raphael.ribeiro@emory.edu}
\affiliation{Department of Chemistry and Cherry Emerson Center for Scientific Computation, Emory University, Atlanta, GA, 30322}
\date{\today}

\begin{abstract}
We introduce a theory of chemical equilibrium in optical microcavities, which allows us to relate equilibrium reaction quotients in different electromagnetic environments. Our theory shows that in planar microcavities under strong coupling with polyatomic molecules, hybrid modes formed between all dipole-active vibrations and cavity resonances contribute to polariton-assisted chemical equilibrium shifts. To illustrate key aspects of our formalism, we explore a model S$_{\t{N}}$2 reaction within a single-mode infrared resonator. Our findings reveal that chemical equilibria can be shifted in either direction of a chemical reaction, depending on the oscillator strength and frequencies of reactant and product normal-modes. Polariton-induced zero-point energy changes provide the dominant contributions, though the effects in single-mode cavities tend to diminish quickly as the temperature and number of molecules increase. Our approach is valid in generic electromagnetic environments and paves the way for understanding and controlling chemical equilibria with microcavities. 
\end{abstract}
\maketitle

\section{Introduction}
\par
Light-matter interactions are almost always irrelevant in equilibrium thermodynamics \cite{mcquarrie2000statistical, pathria2016statistical}. However, recent experiments have suggested otherwise, that the chemical equilibrium of aromatic-halogen charge-transfer complexes may be significantly changed via strong light-matter coupling \cite{equilibrium_1}. 
\par
\begin{figure}
    \centering
    \includegraphics[width=0.25\textwidth]{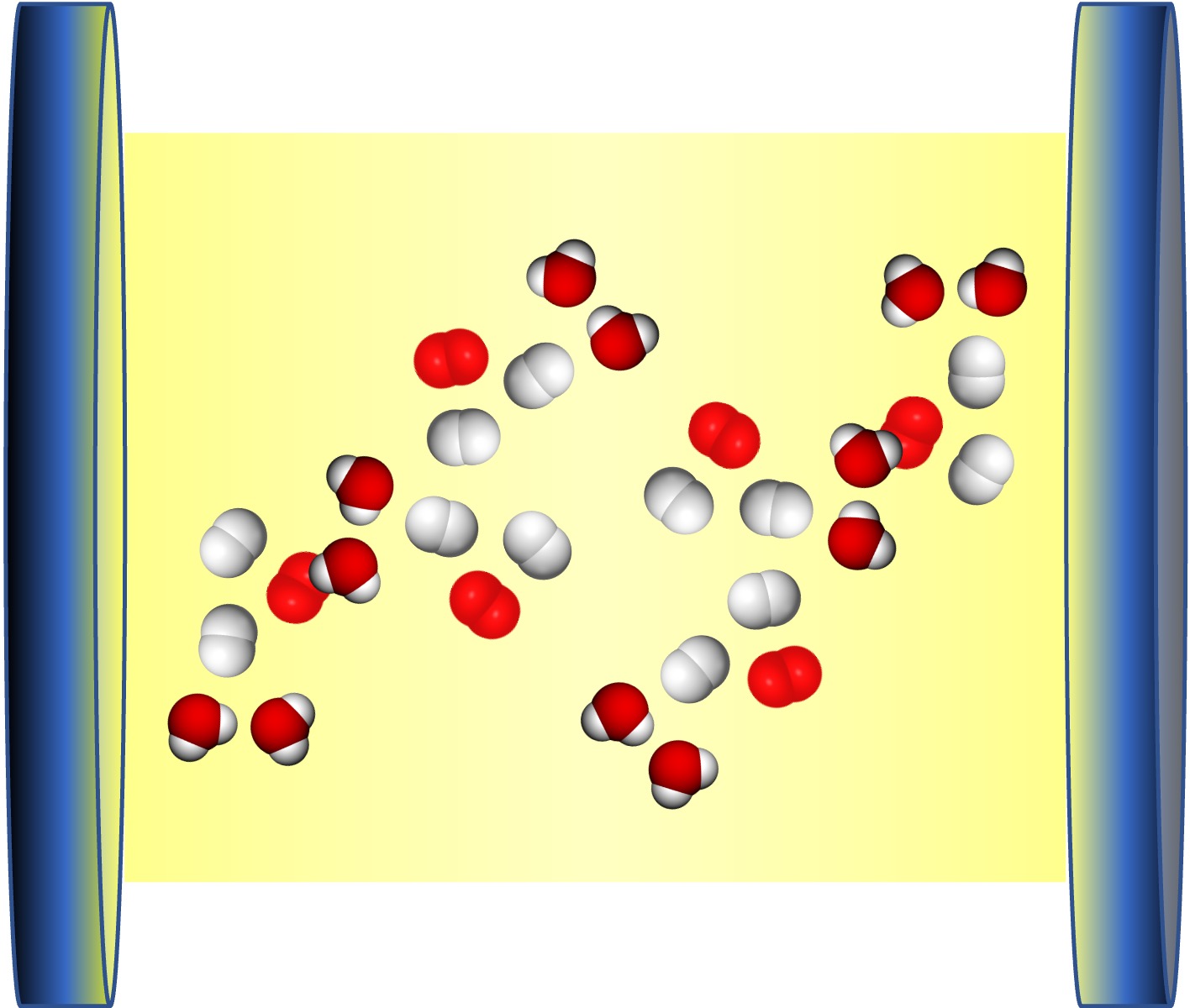}
    \caption{Schematic representation of a reactive mixture in an infrared microcavity supporting confined electromagnetic field modes and strong light-matter interactions, which lead to the formation of hybrid polariton normal-modes with distinct spectra relative to the molecular system in free space and the empty microcavity.}
    \label{fig:cavity}
\end{figure}

 The signature of strong light-matter interactions is the formation of hybrid states referred to as polaritons, consisting of a superposition of electromagnetic (EM) and matter excitations \cite{ebbesen2016hybrid, general_theory_2, hertzog2019strong_LM_interaction_review}. Devices that confine the EM field to the scale of relevant wavelengths [e.g., for infrared (IR) strong coupling, planar cavities are generally constructed with moderate quality mirrors separated by a distance of O($\mu$m)] \cite{kavokin2017microcavities,vahala2003optical_microcavity,bellessa2004strong_plasmon_1,hakala2009vacuum_plasmon_2,chikkaraddy2016single_SC} are generally conducive to polariton formation in the presence of a resonant material (Fig. \ref{fig:cavity}). A simple paradigmatic model of this phenomenon includes an isolated cavity mode under strong interaction with the collective polarization of a molecular system containing $N$ identical  molecules. This system has two hybrid light-matter modes denoted lower and upper polaritons (LP and UP, respectively), and $N-1$ molecular reservoir modes with zero photonic content. 
\par Recent experimental reports have provided evidence that chemical reactions can be substantially affected by strong interactions between IR microcavities and near-resonant molecular vibrational normal modes (vibrational strong coupling) \cite{thomas2016ground, lather2019cavity_reaction_rate, thomas2019tilting, hirai2020modulation,thomas2020ground,lather2020improving_enzyme_cat, sau2021modifying_reaction,sandeep2022manipulating_self_assembly}. Modulation of charge conductivity \cite{conductivity_1,conductivity_2, fukushima2022inherent_conductivity_modification}, and energy transport phenomena \cite{energy_transfer_1,energy_transfer_2,energy_transfer_3,energy_transfer_4,georgiou2018control_energy_transfer} have also been reported. 
\par While theoretical investigations have proposed hypothetical mechanisms for microcavity effects on reaction rates via nonequilibrium effects \cite{theory_reactivity_1, theory_reactivity_2, theory_reactivity_3, theory_reactivity_4, theory_reactivity_5, yang2021quantum_effect_VSC_rates}, less attention has been paid to polariton effects on thermodynamic quantities of molecular systems. Scholes et al. \cite{scholes2020entropy} showed the free energy of dark modes is lower than the polaritonic, and Li et al. \cite{li2020origin} employed classical statistical mechanics to argue that collective strong light-matter coupling is unlikely to affect molecular potentials of mean force. However, recent quantum approaches have shown that polariton effects on thermodynamic quantities could be significant,  especially under ultrastrong coupling conditions \cite{pilar2020thermodynamics_of_USC_System,ashida2020quantum_phase_transition}. 

\par In this work, we present a quantum theoretical investigation of chemical equilibrium under vibrational strong coupling (VSC). We provide a general theory of nonperturbative light-matter interaction effects on chemical equilibrium and obtain an exact relationship between equilibrium reactive mixtures inside and outside microcavities in Sec. \ref{sec:theory}. In Sec. \ref{sec:sn2_example}, we apply our theory to a multicomponent reactive mixture in a single-mode IR cavity resonant with a bright normal-mode of reactants or products. We examine the temperature, normal mode frequency, oscillator strength, and size dependence of the polariton effect on the reactive mixture composition at equilibrium.  We summarize our main results and explain how the provided formalism informs future work in Sec. \ref{sec:conclusions}. 
\section{Theory}\label{sec:theory}
\par 
In this section, we present a general formalism for the investigation of nonperturbative light-matter interaction effects on the composition of generic reactive mixtures. Let A, B, C, and E denote reactive chemical species in equilibrium according to
\begin{align}
	\nu_\t{A} \text{A}+\nu_\t{B}\text{B}\rightleftharpoons \nu_\t{C}\text{C}+\nu_\t{E} \text{E} \label{eq:gen_eq}.
\end{align}
We write the total molecular quantum electrodynamic Hamiltonian \cite{craig1998molecular} for this system as
\begin{align}
H = H_{\t{M}} + H_{\t{L}} + H_{\t{LM}},
\end{align}
where $H_{\t{L}}$ is the transverse EM field Hamiltonian, $H_{\t{M}}$ is the pure matter part of the Hamiltonian  
\begin{align}
H_{\t{M}}=  h_{\t{M}} + V_\t{M}, 	
\end{align}
including the electrostatic intermolecular interactions  in $V_{\t{M}}$, and $h_{\t{M}}$ is the noninteracting (ideal-gas) Hamiltonian for the multicomponent mixture 
\begin{align}
h_{\t{M}} = h_\t{A} + h_{\t{B}} + h_{\t{C}} + h_{\t{E}},
\end{align}
where $h_\t{F}$ corresponds to the Hamiltonian describing a pure ensemble of noninteracting molecules of type $F \in \{\t{A, B, C, E}\}$.
\par We will assume a Born-Oppenheimer description of the molecular system where each molecule is in the electronic ground-state and only vibrational excitations contribute to the light-matter interaction Hamiltonian $H_{\text{LM}}$ representing the photon-matter coupling. 
\par The bare EM field Hamiltonian is given by
\begin{align}
H_{\t{L}} = \sum_{\mbf{k}\lambda}^{k<k_\t{M}} \hbar \omega_{\mbf{k}}\l(a_{\mbf{k}\lambda}^\dagger a_{\mbf{k}\lambda} + \frac{1}{2}\r) + G_{\t{L}},	
\end{align}
where $\mbf{k} = (k_x, k_y, k_z)$, $k = |\mbf{k}|$, $\lambda = 1,2$ denotes the field polarization, and $k_\t{M}$ is a wave vector cutoff (upper bound) for photon modes that will be treated in a nonperturbative fashion below. Only photon modes with $k < k_\t{M}$ are assumed to form polaritons. In a practical application, the value of $k_\t{M}$ would depend on the particular molecular system considered and the strength of the collective light-matter interactions. The contribution of the modes with $k > k_\t{M}$ to $H_{\t{L}}$ is given by the term $G_{\t{L}}$ which we single out since it is not going to provide any contributions to strong light-matter induced changes to chemical equilibria discussed below. 
 \par Vibrational, rotational and translational motions of each molecule are assumed to be separable, and the dipole-active molecular modes undergo small oscillations around equilibrium. For the sake of simplicity, the molecular infrared polarization operator is assumed to be linear with respect to displacements from the equilibrium geometry of each (isolated) chemical species. This is a reasonable assumption under moderate temperatures, especially when strong light-matter coupling happens between IR cavity modes and molecular vibrations with high vibrational temperature $h\nu/k_{\t{B}}$, where $k_{\t{B}}$ is the Boltzmann constant. These approximations lead to a quadratic light-matter Hamiltonian $h_M + H_\t{L} + V_{\t{M}}$ with three types of eigenmodes: (i) polaritons with frequency $\omega_{\alpha}$, (ii) highly off-resonant photon modes, and (iii) molecular translations, rotations, dipole-inactive vibrations described by the Hamiltonian $H_{\t{D}}$ which also includes the ground-state electronic energy contribution of each chemical species. The total light-matter Hamiltonian is given in the new basis by
  \begin{align}
	H &= H_{\t{Pol}}+H_{\t{D}}+G_{\t{L}} + V_{\t{M}},\end{align}
where $H_{\t{Pol}}$ represents the vibrational polariton Hamiltonian in the normal-mode representation
\begin{align}
H_{\t{Pol}} = \sum_{\alpha} \hbar\omega_\alpha\l(c_\alpha^\dagger c_\alpha + \frac{1}{2} \r),
\end{align}
and $c_\alpha^\dagger$ is the bosonic creation operator associated with polariton mode $\alpha$. Note that, in principle, $V_{\t{M}}$ includes all direct intermolecular interactions involving molecular contributions from both $H_{\t{D}}$ and $H_{\t{Pol}}$. As usual, we will assume the interactions modeled by $V_\t{M}$ are weak enough that they can be ignored for the purpose of computing the thermodynamic equilibrium properties of the molecular system. Therefore, we will employ
\begin{align}
h = H_{\t{Pol}}+H_{\t{D}}+G_{\t{L}}	
\end{align}
 to compute the partition function of the light-matter system.
 
 \par The canonical partition function for the system at fixed volume $V$, temperature $T$, including $N_\t{A}, N_\t{B},N_\t{C}, N_\t{E}$ molecules is given by
\begin{align}
Q(N, V, T) & = \t{Tr}~e^{-\beta h} \nonum 
& = Q_{\t{Pol}}(N, V, T) Q_{\t{D}} (N, V, T) Q_{G_{\t{L}}}(V,T),
\end{align}
where $N = (N_\t{A},N_\t{B}, N_\t{C}, N_\t{E})$ and we assumed the electromagnetic field satisfies only fixed volume and temperature constraints (with zero chemical potential since their particle number is not conserved \cite{mcquarrie2000statistical, pathria2016statistical}). The partition function for the electronic-translational-rotational and non-dipole active vibrational degrees of freedom of the molecular system is given by:
\begin{align}
Q_{\t{D}} (N, V, T) = \prod_{\t{F}} \frac{\l[q_{\t{el}}^{\t{F}}(T)q_\t{trans}^{\t{F}}(V,T) q_{\t{rot}}^{\t{F}}(T) \tilde{q}_{\t{vib}}^{\t{F}}(T) \r]^{N_{\t{F}}}}{N_\t{F}!}, \label{eq:qdnvt}
\end{align}
where $\tilde{q}_{\t{vib}}^{\t{F}}(T)$ is the partition function associated with the dipole-inactive normal modes of the F species (single-molecule). It follows that the Helmholtz free energy of the system is
\begin{align}
A(N,V,T) & = -\kb T~\text{ln}[Q(N,V,T)], \nonumber \\ 
&= A_{\t{Pol}}(N,V,T)+ A_{\t{D}}(N,V,T) + A_{\t{G}}(V,T),
\end{align}
where $A_{\t{Pol}}(N,V,T) =-\kb T~\t{ln}[Q_{\t{Pol}}(N,V,T)]$ with $Q_{\t{Pol}}(N,V,T) = \t{Tr}\l[\t{exp}\l(-\beta H_\t{Pol}\r)\r]$ and corresponding expressions apply for $A_{\t{D}}(N,V,T)$ and $A_{\t{G}}(V,T)$.

\par The chemical equilibrium condition at fixed $V$ and $T$ is \cite{mcquarrie2000statistical, pathria2016statistical}
\begin{align}
	\sum_{\t{F}} \nu_\t{F}\frac{\partial{A}}{\partial N_\t{F}}& = -\nu_{\t{A}} \mu_{\t{A}} -\nu_{\t{B}}\mu_{\t{B}} + \nu_{\t{C}} \mu_{\t{C}} + \nu_{\t{E}} \mu_{\t{E}}\nonum 
	& = 0, \label{eq:chemeq_cond}
\end{align}
where the chemical potential of each species is given by
\begin{align}
\mu_\t{F}(N,V,T) = \mu_{\t{F,Pol}}(N,V,T) + \mu_{\t{F,D}}(N_F,V,T), \label{eq:muf_int}\end{align}
with $\mu_{\t{F,Pol}}(N,V,T) = \partial A_{\t{Pol}}/\partial N_\t{F}$ and $\mu_{\t{F,D}}(N_\t{F},V,T) = \partial A_{\t{D}}/\partial N_\t{F}$. We can rewrite Eq. \ref{eq:muf_int} in terms of a bare contribution and a polariton-induced change by adding and subtracting the contribution to the chemical potential from the bright vibrational part of $h_{\t{F}}$ which we write as $\mu_{\t{F,vib, bright}}(T)$.  Noting that $\mu_{\t{F,D}} (N_\t{F},V,T) +\mu_{\t{F,vib,bright}}(T)$ is the bare chemical potential for a system of noninteracting $N_\t{F}$ molecules of type $\t{F}$, we define the bare (for our purposes, the standard-state) chemical potential of species $\t{F}$ by
\begin{align}
	 \mu_{\t{F}}^{(0)}(N_\t{F},V,T) = \mu_{\t{F,D}} (N_\t{F},V,T) +\mu_{\t{F,vib,bright}}(T). \label{eq:muf0}
\end{align}
It follows that the chemical potential of species F under the influence of the light-matter interaction is given by
\begin{align}
	\mu_\t{F}(N,V,T) & = \mu_{\t{F}}^{(0)}(N_F,V,T)+\Delta \mu_{\t{F,Pol}}(N,V,T), \label{eq:muf}
\end{align}
where we introduced 
\begin{align} \Delta \mu_{\t{F,Pol}} = \mu_{\t{F,Pol}}- \mu_{\t{F,vib, bright}}. \label{eq:deltamufpol}
\end{align}

The bare $\t{F}$ chemical potential can be obtained directly from Eqs. \ref{eq:muf0} and  \ref{eq:qdnvt} as
\begin{align}
	\mu_{\t{F}}^{(0)}(N_F, V,T) & = -\kb T\frac{\partial}{\partial N_{\t{F}}}\t{ln}\l[\frac{q_{\t{F}}^{N_\t{F}}(V,T)}{N_{\t{F}}!} \r] \nonum 
	& = -\kb T~\t{ln}\l[\frac{q_{\t{F}}(V,T)}{N_{\t{F}}}\r]. \label{eq:sm_chem_pot}
\end{align}
where $q_{\t{F}}(V,T)$ is the single-molecule partition function for bare isolated species $F$. Inserting Eq. \ref{eq:sm_chem_pot} into Eq. \ref{eq:muf} and using the result in Eq. \ref{eq:chemeq_cond}, we obtain
\begin{align}
	 -\kb T\sum_\t{F} \nu_\t{F} ~\t{ln}\l[\frac{q_\t{F}}{N_{\t{F}}}\r] + \sum_\t{F}\nu_\t{F} \Delta \mu_{\t{F,Pol}} = 0.
\end{align}
A simple rearrangement leads to our expression for the equilibrium reaction (Eq. \ref{eq:chemeq_gen}) quotient under the influence of nonperturbative light-matter interactions 
\begin{align}
	\frac{N_{\t{E}}^{\nu_\t{E}} N_{\t{C}}^{\nu_\t{C}}}{N_{\t{A}}^{\nu_{\t{A}}}N_{\t{B}}^{\nu_\t{B}}} = 	\frac{q_{\t{E}}^{\nu_\t{E}} q_{\t{C}}^{\nu_\t{C}}}{q_{\t{A}}^{\nu_{\t{A}}}q_{\t{B}}^{\nu_\t{B}}} \t{exp}\l[-\beta\sum_\t{F}\nu_\t{F} \Delta \mu_{\t{F,Pol}}(N,V,T)\r]. \label{eq:chemeq_gen}
\end{align}
By solving Eq. \ref{eq:chemeq_gen} for the number of molecules of each species, we obtain the polariton effect on the equilibrium composition of the reactive mixture. 

\par The relation expressed by Eq. \ref{eq:chemeq_gen} indicates several important features of polariton effects on chemical equilibria. For instance, Eq. \ref{eq:chemeq_gen} highlights that a multimode description of the electromagnetic field is essential for modeling the nonperturbative effects of planar microcavities on chemical equilibria, for the polariton contributions to the chemical potential of each component depends on all cavity modes forming polaritons with the considered species. In particular, polaritons originating from all bands of a microcavity in resonance or sufficiently close to resonance with dipole-active molecular vibrations will contribute to Eq. \ref{eq:chemeq_gen}. Clearly, no a priori special role is played by modes corresponding to incidence angles near zero.

\par Equation \ref{eq:chemeq_gen} also demonstrates that in a polyatomic system with multiple bright vibrations, the chemical equilibrium shift induced by an IR microcavity depends on the density of EM modes at the various bright IR resonances of \textit{both} reactants and products and their corresponding oscillator strengths.

\par In the next section, we will apply our equilibrium condition (Eq. \ref{eq:chemeq_gen}) to a reactive mixture where a single molecular species strongly interacts with a microcavity represented by a single mode. For this purpose, let us consider here the following example. Suppose only the reactant species $\t{A}$ strongly interacts with the single-mode cavity EM field, and the mean equilibrium number of molecules of type A (obtained from solving Eq. \ref{eq:chemeq_gen}) is $N_\t{A}$. It follows the nonperturbative light-matter Hamiltonian contains $N_\t{A}+1$ eigenmodes corresponding to the $N_\t{A}-1$ purely molecular modes that have the same spectrum as the bright vibrations of $\t{A}$ and the hybrid LP and UP. The contribution of the $N_\t{A}-1$ reservoir normal modes to $\mu_{\t{F,Pol}}$ cancels out the term $\mu_{\t{A,vib,bright}}(T)$ in $\Delta \mu_{\t{A,Pol}}(N,V,T)$ (Eq. \ref{eq:deltamufpol}). As expected, the effect of nonperturbative light-matter interactions on the composition of the reactive mixture at equilibrium is entirely due to the LP and UP modes. We determine the polariton effect on the chemical equilibrium $F_{\t{Pol}}(V,T)$ by computing the ratio of the equilibrium reaction quotient inside the microcavity $R(V,T) = N_\t{E}^{\nu_\t{E}} N_\t{C}^{\nu_\t{C}}/N_\t{A}^{\nu_\t{A}} N_\t{B}^{\nu_\t{B}}$ to the standard-state reaction quotient (equilibrium constant) $K^{(0)}(T) = (q_\t{E}^{\nu_\t{E}}q_\t{C}^{\nu_\t{C}})/(q_\t{A}^{\nu_\t{A}}q_\t{B}^{\nu_\t{B}})$. It follows from Eq. \ref{eq:chemeq_gen} that 
\begin{align}
	F_{\t{Pol}}(V,T) & = \frac{R(V,T)}{K^{(0)}(T)} \nonum 
	& = e^{\beta \nu_ \t{A}
\l[\mu_{\t{A}}^{\t{LP}}(N_A,V,T) +\mu_{\t{A}}^{\t{UP}}(N_A,V,T)\r]}, \label{eq:fpol_single}\end{align}
where the LP and UP contributions to the chemical potential of the strongly coupled chemical species $\t{A}$ is given by 
\begin{align}
	\mu_{\t{A}}^{\t{LP}}(N_\t{A},V,T) = \frac{\partial A_{\t{LP}}(N_\t{A}, V, T)}{\partial N_\t{A}},
\end{align}
 and $A_{\t{LP}} = -\kb T~\t{ln} ~q_{\t{LP}}(N_\t{A}, V,T)$, and identical definitions exist for UP.

\section{Bimolecular nucleophilic substitution reaction in a single-mode cavity}\label{sec:sn2_example}
\subsection{Model}
To illustrate the theory described above, we consider a lossless cavity interacting with a gas-phase reactive mixture where equilibrium is established via the S$_\t{N}$2 reaction
\begin{equation}
\text{C}_{2}\text{H}_{5}\text{Br}+\text{Cl}^{-} \rightleftharpoons\text{C}_{2}\text{H}_{5}\text{Cl}+\text{Br}^{-}. \label{eq:sn2_eq}
\end{equation}
This reaction has been thoroughly studied in the gas phase \cite{li1996high_gas_phase_Sn2_experimental,hase1994simulations_gas_phase_Sn2_simulation}. We construct its chemical equilibrium constant in free space from the gas-phase partition function of each chemical species assuming separability between the internal degrees of freedom and ideal gas conditions. We ignore the electronic contribution to the reaction free energy, for in this case, the equilibrium constant is close to 1 at room temperature and strong light-matter coupling becomes more likely to significantly affect the composition of the reactive mixture at equilibrium. Vibrational contributions to the partition function were constructed using the quantum harmonic oscillator model, whereas classical partition functions were employed in the treatment of rotational and translational degrees of freedom of each species. Vibrational frequencies and geometric parameters were extracted from data available in the Chemistry WebBook \cite{reaction_data}. 

\par In order to probe polariton effects on the chemical equilibrium associated with Eq. \ref{eq:sn2_eq}, we suppose the system is embedded in an optical cavity with a single high-quality mode in resonance with a single vibrational mode of reactants or products. To examine the distinct effects of reactant and product strong light-matter coupling, we chose two strongly absorbing IR modes of reactants and products in the gas phase \cite{reaction_data}. The frequencies of the selected dipole-active normal modes are given in Table \ref{tab:modes}.

\par When a single normal mode (of $N_{\t{F}}$ reactant or product molecules) interacts nonperturbatively with the optical microcavity, the polaritonic part of the Coulomb gauge \cite{craig1998molecular} light-matter Hamiltonian can be written in the uncoupled basis as \cite{kavokin2017microcavities}
\begin{align}\label{eq:single_mode_hamiltonian}
\begin{split}
H=&\sum_{i=1}^{N_{\t{F}}}\hbar\omega_\t{F}a^{\dagger}_{i}a_{i}+\hbar\tilde{\omega}_{\t{C}}b^{\dagger}b ~+\\
&g\sqrt{\frac{\omega_\t{M}}{\tilde{\omega}_\t{C}}}\sum_{j=1}^{N_{\t{F}}}(a^{\dagger}_{j}-a_{j})(b^{\dagger}+b),
\end{split}
\end{align}
 where $\omega_{\t{F}}$ is the frequency of the strongly coupled molecular normal mode  (of type $\t{F}$), $a_{i}^{\dagger}$ and $a_{i}$ are the creation and annihilation operators of $\t{F}$ excitations in the $i$th molecule, and $b^{\dagger}$ and $b$ are the creation and annihilation operators of the cavity mode with renormalized frequency $\tilde{\omega}_{\text{C}}$ given by
\begin{align}\tilde{\omega}^{2}_{\text{C}}&=\omega^{2}_{\text{C}}+\Omega^{2}_{\text{R}}, \label{eq:cav_renorm}
\end{align}
 where $\omega_{\text{C}}$ is the bare photon frequency and $\Omega_{\t{R}} = 2g\sqrt{N_\t{F}}$ is the collective light-matter interaction strength. We have assumed throughout that the bare cavity mode is on resonance with a reactant or product normal mode (Table 1), therefore, we set $\omega_{\t{C}} = \omega_\t{F}$ from now on.  When the renormalized cavity mode frequency is near-resonant with the molecular normal mode, the effective collective light-matter interaction strength for the strongly coupled species is  $g\sqrt{N \omega_{\t{F}}/\tilde{\omega}_\t{C}} \approx g\sqrt{N_\t{F}}$. Note that the reactant and product modes listed in Table \ref{tab:modes} have essentially equal oscillator strength \cite{reaction_data}, and, therefore we will employ the same value of $g$ when analyzing the effects on chemical equilibrium induced by exclusive strong light-matter coupling with each mode.
 
 \par The light-matter system described by the Hamiltonian given by Eq. \ref{eq:single_mode_hamiltonian} has $N_{\t{F}}+1$ eigenmodes. The  frequencies of the hybrid excitations (polaritons) are
\begin{align}
    \omega_{\t{LP}} &=\sqrt{\frac{\tilde{\omega}^{2}_{\text{C}}+\omega^{2}_\text{F}-\sqrt{(\tilde{\omega}^{2}_{\text{C}}-\omega^{2}_\text{F})^{2}+4\Omega^{2}_{\text{R}}\omega^{2}_\text{F}}}{2}},\\
    \omega_{\t{UP}} &=\sqrt{\frac{\tilde{\omega}^{2}_{\text{C}}+\omega^{2}_\text{F}+\sqrt{(\tilde{\omega}^{2}_{\text{C}}-\omega^{2}_\text{F})^{2}+4\Omega^{2}_{\text{R}}\omega^{2}_\text{F}}}{2}},
\end{align}
whereas the remaining $N_{\t{F}}-1$ reservoir modes have the same frequency $\omega_\t{F}$ as the bare molecules.

\par Using basic statistical mechanics \cite{mcquarrie2000statistical, pathria2016statistical}, we can obtain the polariton and reservoir mode partition functions and compute the polariton effect on the chemical potential $\Delta\mu_{\t{F,Pol}}$ (Eq:\ref{eq:deltamufpol}) required to set up the nonlinear Eq. \ref{eq:chemeq_gen}. Its solution consists of the equilibrium number of molecules of each species inside the optical cavity, and allows us to establish the polariton effect on the chemical equilibrium as measured by $F_{\t{Pol}}(V,T)$ via Eq. \ref{eq:fpol_single}.

The numerical problem is set up by assuming that the mixture initially contains an equal number of ethyl bromide and chloride ions $N = N_{\t{C}_2\t{H}_5\t{Br}}^0 = N_{\t{Cl}^-}^0$ that react according to Eq. \ref{eq:sn2_eq} to establish equilibrium with ethyl chloride and bromine ions. The number of reactant and product molecules at equilibrium is denoted $N_\t{R}$ and $N_\t{P}$ respectively. It follows that at equilibrium $N_\t{R}=  N_{\t{C}_2\t{H}_5\t{Br}} = N_{\t{Cl}^-}$ and $N_\t{P} = N_{\t{C}_2\t{H}_5\t{Cl}} = N_{\t{Br}^-} = N - N_\t{R}$. The standard-state equilibrium constant $K_0(T)$ (outside the microcavity) is computed as a function of temperature using the ratio of product and reactant partition functions. To find the equilibrium composition of the mixture at thermal equilibrium, we solve the equation  
\begin{align}
\frac{(N - N_\t{R})^2}{N_\t{R}^2}	 = K_0(T) e^{\beta \nu_ \t{F}
\l[\mu_{\t{F}}^{\t{LP}}(N_\t{F},V,T) +\mu_{\t{F}}^{\t{UP}}(N_\t{F},V,T)\r]}, \label{eq:selfconsistent}
\end{align}
for $N_\t{R}$, where $\nu_{\t{F}}$ is the stoichiometric coefficient of the strongly coupled species, and $N_{\t{F}} = N_{\t{R}}$ or $N_{\t{P}}$ when strong coupling occurs with a reactant or product normal mode, respectively. We solve Eq. \ref{eq:selfconsistent} for a given $T$, initial number of molecules $N$, and single-molecule light-matter interaction strength $g$. The standard-state equilibrium composition of the reactive mixture is employed as an initial guess for the solution, and the polariton contributions to the chemical potential are obtained from automatic differentiation of the polariton free energies with respect to the number of strongly coupled molecules as implemented in the  python AutoGrad package \cite{maclaurin2015autograd}. 

In the case where a reactant mode strongly interacts with the microcavity, the polariton effect on the equilibrium composition of the reactive mixture can be written as a ratio of the equilibrium constant inside the cavity to the bare equilibrium constant (Eq. \ref{eq:chemeq_cond})
\begin{align}
F_{\t{Pol}}^{\t{R}}&=\t{exp}\l[-\beta \mu_{\t{R},\t{Pol}}\l(N_{\t{R}}\r)\r],	
\end{align}
where $\mu_{\t{R,Pol}}(N_\t{R}) = \mu_{\t{R}}^{\t{LP}}(N_\t{R}) + \mu_{\t{R}}^{\t{UP}}(N_\t{R})$. Conversely, if strong coupling occurs with the product molecules, we obtain
\begin{align}F_{\t{Pol}}^{\t{P}}&=\t{exp}\l[\beta\mu_{\t{P,Pol}}\l(N_{\text{P}}\r)\r].
\end{align}
\renewcommand{\arraystretch}{1.25}
\begin{table}

\begin{tabular}{ |c|c|c|c| }
\hline
$\text{C}_{2}\text{H}_{5}\text{Cl}$ mode & $\omega \l(\t{cm}^{-1}\r)$  &$\text{C}_{2}\text{H}_{5}\text{Br}$ mode &  $\omega \l(\t{cm}^{-1}\r)$ \\ 
\hline
CC Str & 974 & CC Str & 964 \\
CCl Str & 677& CBr Str & 583  \\
\hline
\end{tabular}
\caption{\label{tab:modes} Selected IR-active vibrational modes of ethyl chloride and ethyl bromide with frequencies obtained from Ref. \cite{NIST_data}.}
\end{table}

\subsection{Results and Discussion}
\par We have investigated the effect of single-mode strong light-matter coupling on the equilibrium composition of the reactive molecular mixture described by Eq. \ref{eq:sn2_eq} at various temperatures, system sizes, and light-matter interaction strengths assuming that strong coupling occurs between the cavity and a single set of normal vibrational modes of reactant or product. 

\par The bare cavity frequency $\omega_\t{C}$ is set to be on resonance with the strongly coupled vibrational mode. Renormalization (Eq. \ref{eq:cav_renorm}) of the cavity frequency in the presence of the molecular system leads to a nonzero detuning that is insignificant relative to the light-matter interaction strength under the conditions analyzed in this work. 

\par
\begin{figure}[ht]
    \centering
    \includegraphics[width=0.45\textwidth]{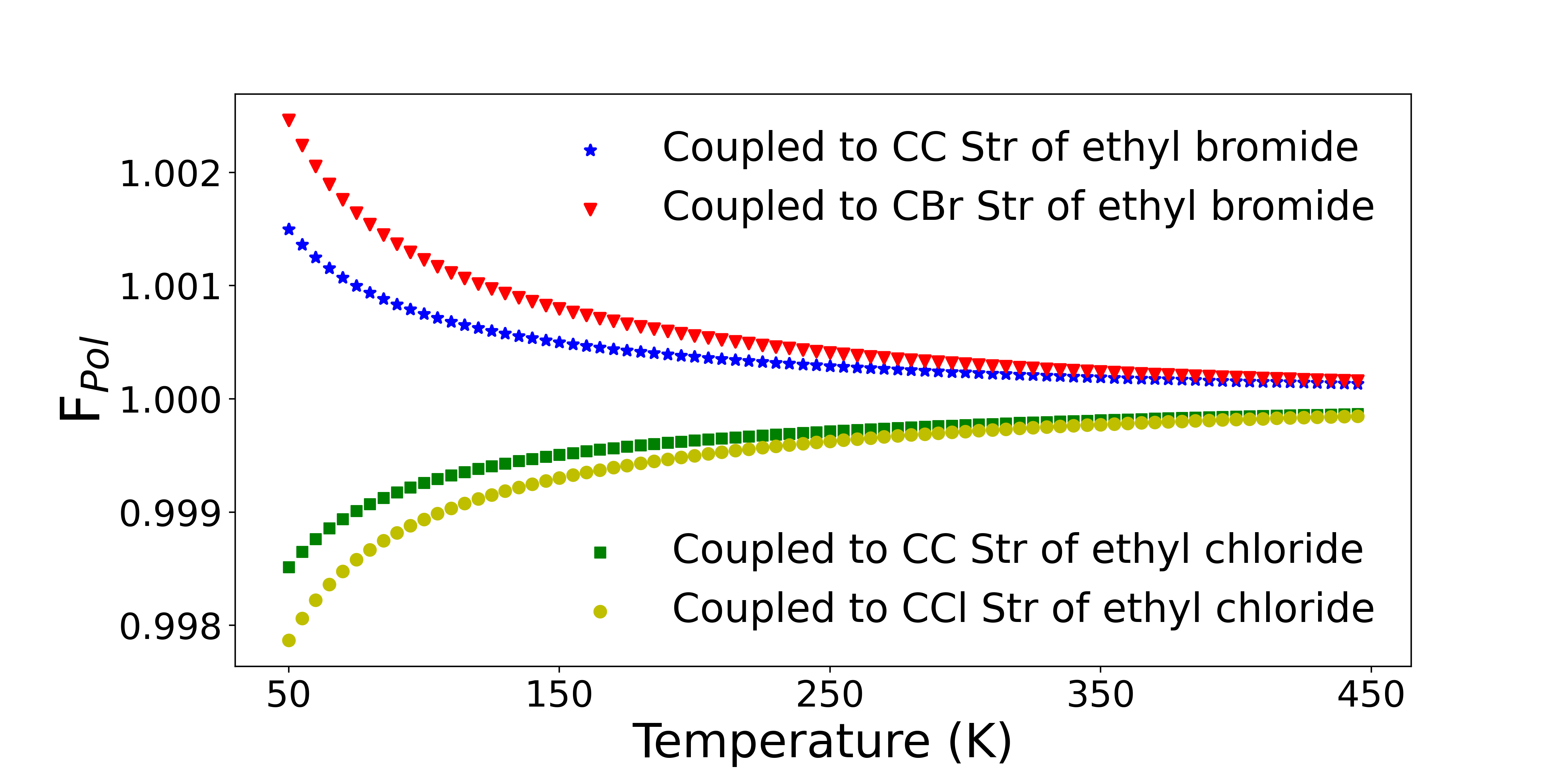}
    \caption{Temperature dependence of the single-mode cavity effect on the examined S$_\t{N}$2 equilibrium. The single-molecule light-matter coupling strength is $g = 10~\text{cm}^{-1}$ and the maximum number of strongly coupled modes is $N = 100$.}
    \label{fig:temp_dependence}
\end{figure}

 The temperature dependence of the ratio between the reaction quotient of the selected S$_\t{N}$2 reaction inside and outside a microcavity is provided in Fig. \ref{fig:temp_dependence}. This figure shows four notable features: a. polariton effects are strongest at low temperatures and vanish at the high-temperature limit, b. the equilibrium is shifted towards the products when reactants are strongly coupled to light and vice-versa, c. the observed effects are extremely small considering the large single-molecule light-matter coupling strength employed (for the purposes of illustrating our theory), and d. polaritons formed between molecular modes with lower frequency have a stronger impact on the chemical equilibrium constant. Below, we discuss each of these trends.

\begin{figure}
    \centering
    \includegraphics[width=0.45\textwidth]{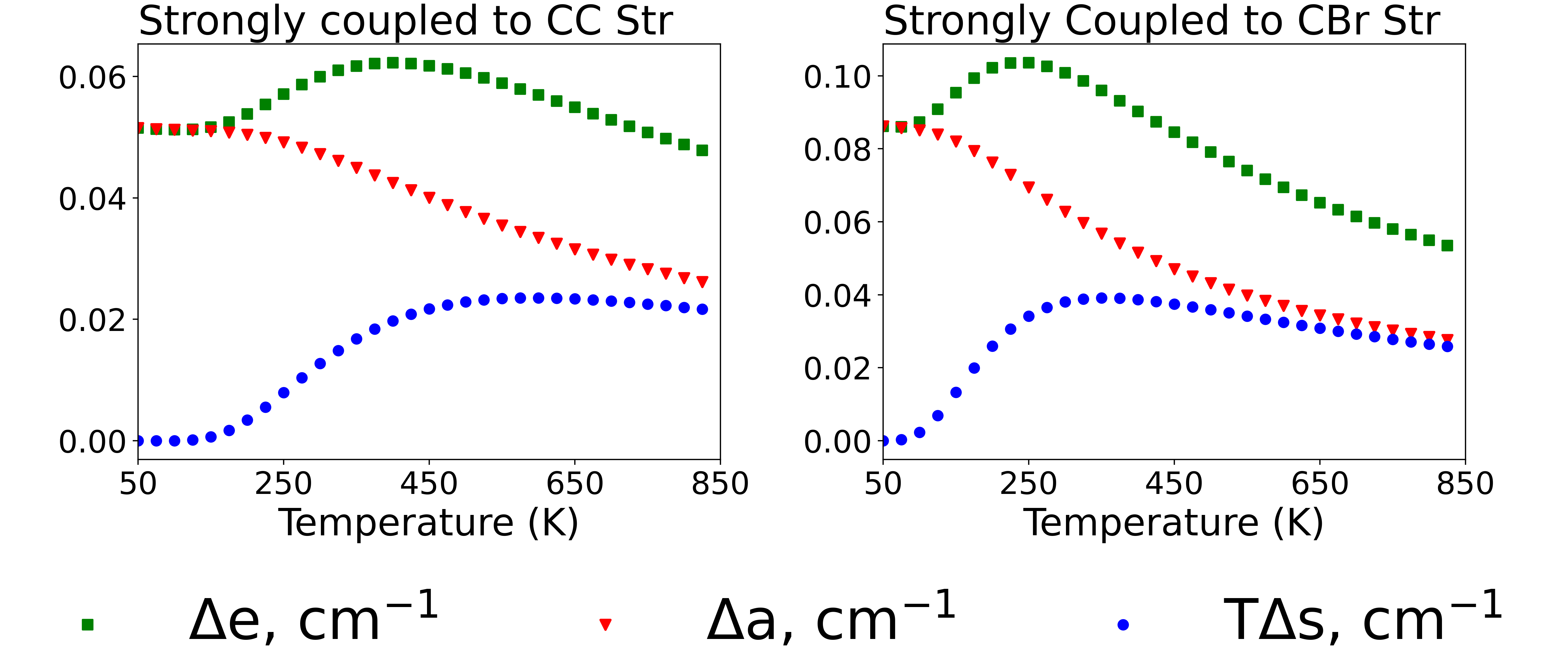}
    \caption{Polariton-induced changes in thermal properties per unit interacting degree of freedom ($\Delta a$, $\Delta e$, and $T\Delta s$) as a function of temperature. The left figure shows the results obtained when the CC Str mode of ethyl bromide is strongly coupled to the cavity, with $g = 10~\text{cm}^{-1}$. The right figure shows analogous results when the CBr Str mode is strongly coupled to the cavity.}
\label{fig:thermodynamic_T_analysis}
\end{figure}

\textit{Low- and high-temperature behavior.} Figure \ref{fig:temp_dependence} shows the single-mode cavity effect on the composition of the equilibrium reactive mixture is largest at low temperatures, whereas strong coupling has no effect in the high-temperature limit. To understand this, note that at low temperatures, the polaritons and bare modes are essentially in their ground-state, and therefore any polariton-induced change in free energy responsible for modifying chemical equilibrium is generated by the difference between polariton and bare molecule zero-point energies. At high temperatures,  the classical limit of the light-matter partition function can be employed to show that the free energy of the reactive mixture is unaffected by polariton formation \cite{li2020origin,pilar2020thermodynamics_of_USC_System}. 

\par In Fig. \ref{fig:thermodynamic_T_analysis}, we examine the polariton-induced change per strongly coupled degree of freedom (molecular and photonic) in the internal energy $\Delta e = \Delta E/(N_\t{F}+1)$, free energy $\Delta a = \Delta A/(N_{\t{F}}+1)$ and $T\Delta s = T\Delta S/(N_{\t{F}}+1)$ of the system at equilibrium
\begin{align}
 &\Delta e =\frac{E_{\text{LP}}+E_{\text{UP}}-E_\t{F}-E_{\t{C}}}{N_\t{F}+1}, \\
 & \Delta a=\frac{A_{\text{LP}}+A_{\text{UP}}-A_\t{F}-A_{\t{C}}}{N_\t{F}+1}, \\
  & T\Delta s=\frac{S_{\text{LP}}+S_{\text{UP}}-S_\t{F}-S_{\t{C}}}{N_\t{F}+1},
\end{align}
where $\t{F}$ is either R or P and $N_{\t{F}}$ is the number of strongly coupled molecules at equilibrium. We limit our discussion to strong coupling with the reactant ensemble (C$_2$H$_5$Br) since the conclusions we derive here are straightforwardly generalizable to the case where strong coupling occurs with products.

\par Fig. \ref{fig:thermodynamic_T_analysis} shows the observed polariton effect in the reactive mixture composition (Fig. \ref{fig:temp_dependence}) at low $T$ is essentially due to the cavity-induced change in reactant or product zero-point energies. This follows from the fact that at the low-$T$ limit, $\Delta e$ is entirely determined by the zero-point energy of the degrees of freedom involved in strong light-matter coupling
\begin{align}
    \lim_{T\rightarrow 0}\Delta{e}&=\frac{\hbar (\omega_{\text{LP}}+\omega_{\text{UP}}-\omega_\t{F}-\omega_{\t{C}})}{2(N_\t{F}+1)}. \label{eq:zpe_lowT}
\end{align}
Conversely, the entropy contribution of all modes vanish as $T \rightarrow 0$. Therefore, it follows, given that $\omega_{\text{LP}}+\omega_{\text{UP}}-\omega_\t{F}-\omega_{\t{C}}\ne 0$, the change in system free energy induced by the optical cavity at low temperatures relative to the vibrational temperature of the strongly coupled modes is dominated by the ground-state energy difference between the polariton normal-modes and the microcavity and molecular vibrational modes. 

 \par At high temperatures, the polariton effect on the composition of the molecular mixture at thermodynamic equilibrium goes to zero regardless of the vibrational frequency and light-matter coupling strength. The absence of any effect on the internal energy may be seen from the equipartition theorem (this implies that each normal mode has $k_{\t{B}}T$ mean internal energy) \cite{mcquarrie2000statistical, pathria2016statistical}, whereas entropy variations induced by strong light-matter coupling may be seen to vanish from the classical limit of the harmonic oscillator partition functions which can be directly applied to give
\begin{align}\label{entropy_high_temp}
\lim_{T\rightarrow \infty}\Delta{s} & \approx \frac{1}{N_\t{F}+1}\text{ln}\left(\frac{q_{\text{LP}} q_{\text{UP}}}{q_{\t{F}}q_{\t{C}}}\right)\nonumber \\
    &=\frac{1}{2(N_{\t{F}}+1)}\text{ln}\left(\frac{\omega^{2}_{\t{C}}}{\omega^{2}_{\t{C}}+\Omega^{2}_{\t{R}}-\Omega^{2}_{\t{R}}}\right) = 0. \end{align} 
It follows that $T \Delta s$ goes to 0 at low and high $T$ but is an increasing function of $T$ at intermediate temperatures, therefore showing a maximum (Fig. \ref{fig:thermodynamic_T_analysis}). 
    
\textit{Direction of chemical equilibrium shift.} Figure \ref{fig:thermodynamic_T_analysis} also explains why single-mode strong coupling with a chemical species tends to bias the equilibrium towards the uncoupled species. This occurs because the sum of polariton zero-point energies $E_\t{LP} + E_{\t{UP}}$ is greater than the sum of the bare molecule normal-mode and bare photon zero-point energies. This feature increases the free energy of the light-matter system inside the microcavity relative to the bare system.
    
 \textit{Magnitude of polariton effect on chemical equilibrium.} The single-mode strong coupling effect on chemical equilibrium as measured by the ratio of the reaction quotient in the microcavity to the standard-state (bare) equilibrium constant is observed to be less than 1.003 even at low temperatures such as 50 K. The effect becomes even weaker at higher temperatures. This may also be understood from Fig. 3, which shows that the (single-cavity mode) polariton formation effect on the free energy change \textit{per degree of freedom} is extremely small. We revisit this point when discussing the system-size dependence of our results later.
 
 \par \textit{Strong coupling with lower frequency normal-modes have greater impact on chemical equilibrium constants}. Fig. \ref{fig:temp_dependence} shows that strong light-matter coupling is most effective (among the scenarios we considered) when the matter part of polaritons corresponds to the CCl (product normal-mode) or CBr (reactant normal-mode) stretch modes. These motions have lower frequency than the CC stretch of either reactants or products by about 300  and 400 cm$^{-1}$, respectively. The greater impact of VSC occurring with lower frequency vibrations may be understood mathematically from an analysis of the polariton contribution to the zero-point energy difference between the polaritonic system and the composite (light-matter) bare system per (strongly coupled) degree of freedom. Under the conditions examined here where $\tilde{\omega}_{\t{C}}\approx\omega_{\t{M}}$ and $g/\omega_M \ll 1$, the polariton effect at the zero-point energy difference is given by
 \begin{align}  \lim_{T\rightarrow 0}\Delta{e} \approx \frac{g^2}{2\omega_\t{M}}, ~~g/\omega_M \ll 1.
\label{eq:ZPElimit}   
 \end{align}
This result clearly demonstrates that at low temperatures, where single-mode cavity effects on equilibrium are largest, light-matter interactions will have more significant impact when it involves lower frequency and stronger oscillator strength vibrational modes \cite{pilar2020thermodynamics_of_USC_System}. Similar results are valid at higher temperatures where thermal excitations play a greater role but ultimately lead to no polariton effect at chemical equilibrium in the $T \rightarrow \infty$ limit \cite{pilar2020thermodynamics_of_USC_System}.

\par
\begin{figure}
    \centering
    \includegraphics[width=0.45\textwidth]{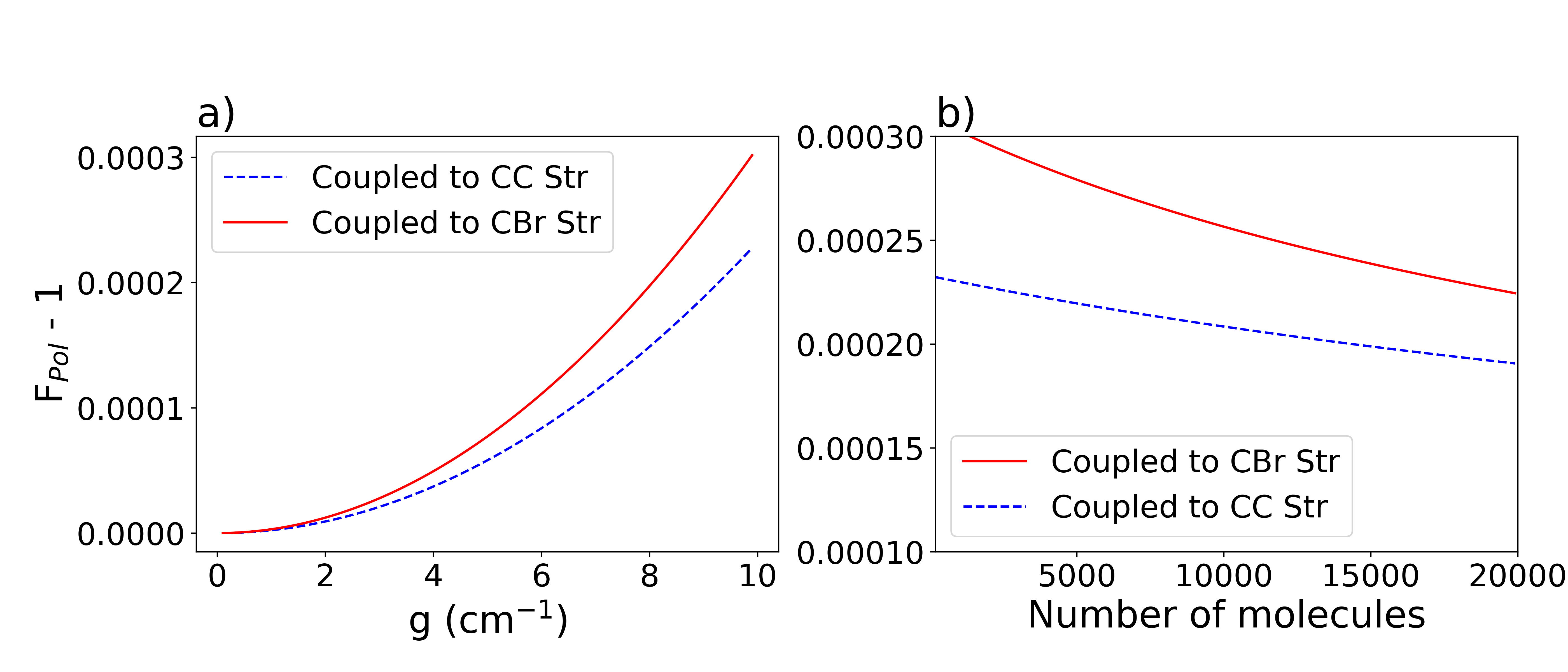}
    \caption{(a) Polariton effect on chemical equilibrium constant vs single-molecule light-matter coupling strength for a system with a maximum number of 100 product molecules at $T  = 300$ K. (b) Polariton effect on equilibrium at $T = 300$ K and $g=10~\text{cm}^{-1}$ vs total number of molecules inside cavity.}
    \label{fig:merge}
\end{figure}

 \textit{Size and oscillator strength dependence of polariton effects on chemical equilibrium.} In the examined reaction, all strongly coupled modes have nearly equal oscillator strength. However, this is not a generic feature of polyatomic molecules, which will generally have vibrational excitations with variable absorption intensity. In order to assess the dependence of $F_{\t{Pol}}$ on the single-molecule light-matter coupling strength, we present in Fig. \ref{fig:merge}(a) the behavior of $F_{\t{Pol}}$ at $T = 300~\t{K}$ as a function of the single-molecule light-matter coupling constant. As expected, the polariton effect on the composition of the reactive mixture is enhanced with increasing single-molecule light-matter coupling strength. This behavior follows expectations based on Eq. \ref{eq:ZPElimit} and others similarly related \cite{pilar2020thermodynamics_of_USC_System}.
 
\par We end our analysis of single-mode microcavity effects on chemical equilibrium by quantitatively investigating the behavior of the polariton effect under changes in the maximum number of strongly coupled molecules $N$ (with fixed cavity volume). We find that while $\Omega_{\t{R}}$ increases, the overall VSC effect on the reactive mixture composition decreases substantially as the number of molecules increases at $T=300\t{K}$ [Fig. \ref{fig:merge}(b)]. 

\par The weakening of VSC-induced changes on chemical equilibrium constants with increasing molecular density is an expected feature of single-cavity mode theories \cite{galego2015cavity,li2020origin, ribeiro2022introduction} which have systematically shown that polariton effects on local molecular observables decrease with increasing molecular density. Fig. \ref{fig:merge} shows a substantial deviation of the scaling of the strong coupling effect on the examined chemical equilibrium constant relative to simple $1/N$ scaling, but the implication of various earlier studies \cite{galego2015cavity, campos2019resonant, zhdanov2020vacuum, li2020origin, campos2020polaritonic} remain valid that single-mode cavity effects on local molecular observables are insignificant in the collective light-matter interaction regime.
  
\section{Conclusions}\label{sec:conclusions}
We provided a general theory of chemical equilibrium under nonperturbative light-matter interactions and applied it to an S$_{\t{N}}$2 reaction in a single-mode cavity. We found that polaritons can shift chemical equilibrium constants towards either direction of a reaction depending on the species (reactant or product) strongly coupled to the EM field, light-matter interaction effects are strongest at lower temperatures, and the change induced by VSC on the internal energy of the light-matter system provide the dominant contribution to the VSC effect on chemical equilibria. We also showed that strong light-matter coupling is more effective at shifting chemical equilibria when polaritons are formed between IR cavity modes and molecular vibrations with larger oscillator strength and lower frequency.

\par These trends were obtained in an idealized scenario where strong light-matter coupling occurs between a single EM mode of an IR resonator and the normal-modes of a particular component of the reactive mixture (reactant or product ensemble) but are based on fundamental features of our theory that are expected to hold more generally. Future work, based on Eq. \ref{eq:chemeq_gen} including a quantitative analysis of equilibrium effects in realistic systems with simultaneous strong coupling of multiple IR modes of reactants and products with multimode Fabry-Perot cavities will allow direct comparison with experiments \cite{equilibrium_1}.

\section{Acknowledgments} 
RFR acknowledges generous startup funds from Emory University.
\section{References}
\bibliography{reference.bib}

\end{document}